\documentclass[proceedings]{JHEP}
\conference{Non-perturbative Quantum Effects 2000}
\usepackage{epsfig}
\def\EQ{\begin{equation}}
\def\EN{\end{equation}}

\title{Universal Ratios of the Renormalization Group}

\author{G. Mussardo \\Universita' dell'Insubria, Como (Italy) \\
INFN, Sezione di Trieste (Italy)} 
\abstract{
The scaling form of the free--energy near a critical point
allows for the definition of various universal ratios of thermodynamical 
amplitudes. Together with the critical exponents they characterize the 
universality classes and may be useful experimental quantities.
We show how these universal quantities can be computed for 
a particular class of universality by using several Quantum Field 
Theory methods}

\begin{document}
\section{Introduction}

%{\large {\bf Introduction}}

\noindent
One of the most important successes of Quantum Field Theory (QFT) 
in recent years consists of the quantitative analysis of the
universality classes of two--dimensional statistical mechanics 
models near their second order phase transition points. Right 
at criticality, where the correlation length $\xi$ diverges, 
Conformal Field Theory (CFT) allows the exact determination of the 
spectrum of anomalous dimensions, the structure constants of 
the OPE algebra, the multi--point correlation functions etc. 
\cite{BPZ}. However, the CFT data do not exhaust all physical 
information relative to the phase transitions. From a theoretical 
point of view, a perturbation of the conformal action is required 
in order to investigate the space of the coupling constants and 
its topology consisting of the location of the fixed points and 
the Renormalization Group flows which connect them. This rich and 
interesting subject has been deeply investigated by several 
authors in the last ten years (see, for instance, \cite{GMrep} and 
references therein for a review). 
From a practical point of view, one has to take into account that 
in a real sample, conformal invariance may be broken by the presence 
of impurities or, simply, by an imperfect fine--tuning of the 
experimental knobs. Hence, to control these effects it becomes 
important to study the dynamics of the systems slightly away 
from criticality, in particular their responses to external fields. 
The most ambitious goal would be the determination of both the 
scaling function which fixes the equation of state and the 
off--critical correlators of the various order parameters. 
Although the exact determination of the equation of state of 
a given universality class may often be a difficult 
task, the scaling property alone of the free--energy is 
nevertheless sufficient to extract numerous predictions on 
universal combinations of critical amplitudes. As will be 
discussed below, these universal combinations are pure numbers 
which can be extremely useful for the experimental identification 
of the universality classes.   

\section{Universal Amplitude Ratios}
%{\large {\bf Universal Amplitude Ratios}}

\noindent
Consider a statistical model with $n$ relevant fields 
$\varphi_i(x)$ at criticality. Near the critical point, 
its action can be parameterised as 
\EQ
{\cal A} = {\cal A}_{CFT} + g_i \int \varphi_i(x) d^Dx
\,\,\,.
\label{actionoff}
\EN 
The scaling property of the order parameters is encoded into
the asymptotic form of their two--point functions
$ 
\langle \varphi_i (x) \varphi_i(0) \rangle
\simeq \frac{1}{\hspace{1mm}|x|^{4 \Delta_i}}
$ 
for $|x|\rightarrow 0$ 
so that $\Delta_i$ is identified with the conformal dimension of 
the fields. Correspondingly the conjugate coupling constants $g_i$ 
behave as
$ 
g_i \sim \Lambda ^{D - 2 \Delta_i} \,\,\, ,
$ where $\Lambda$ is a mass scale. Therefore, away from criticality 
there will be generally a finite correlation length $\xi$ which in the 
thermodynamical limit scales as
$
\xi \sim a \,(K_i g_i)^{-\frac{1}{D - 2 \Delta_i}} \,\,\,,
$
where $a \sim \Lambda^{-1}$ may be regarded as a microscopic
length scale. The terms $K_i$ are non--universal metric factors
which depend on the unit chosen for measuring the external sources
$g_i$, alias on the particular realization selected for representing
the universality class. In the presence of several deformations of
the conformal action, the most general expression for the scaling
form of the correlation length may be written as
\EQ
\xi =\xi_i \equiv a \,(K_i g_i)^
{-\frac{1}{D -2 \Delta_i}}\,
{\cal L}_i\left(\frac{K_j g_j}{(K_i g_i)^{\phi_{ji}}}\right) \,\,\,,
\label{xii}
\EN
where
$ 
\phi_{ji} \equiv \frac{D-2 \Delta_j}{D-2 \Delta_i}
$
are the so--called {\em crossover exponents} whereas ${\cal L}_i$        
are universal homogeneous scaling functions of the ratios
$\frac{K_j g_j}{(K_i g_i)^{\phi_{ji}}}$. There are of course
several (but equivalent) ways of writing these scaling forms,
depending on which coupling constant is selected as a prefactor.
In the limit where $g_l \rightarrow 0$ ($l \neq i$) but
$g_i \neq 0$, equation (\ref{xii}) becomes
\EQ
\xi_i = a \,\xi_i^0 \,g_i
^{-\frac{1}{D -2 \Delta_i}}
\,\,\,\,, \,\,\,\, 
\xi_i^0 \sim K_i^{-\frac{1}{D -2 \Delta_i}} \,\,\,.
\label{xio}
\EN

Consider now the free--energy $\hat f[g_1,\ldots,g_n]$. This is
a dimensionless quantity defined by
\EQ
e^{-\hat f(g_1,\ldots,g_n)} = 
\int {\cal D}\phi \, e^{- \left[
{\cal A}_{CFT} +\sum_{i=1}^n g_i \int \varphi_i(x) d^Dx
\right]} \,\,\,.
\label{free}
\EN
Assuming the validity of the hyperscaling hypothesis, in the
thermodynamical limit its singular part (per unit of volume)
will be proportional to the $D$ power of the correlation length.
Let us denote the singular part of the free--energy for unit
volume by $f[g_1,\ldots,g_n]$. Depending on which scaling form
is adopted for the correlation length, we have correspondingly
several (but equivalent) ways of parameterizing this quantity
\EQ
f_i[g_1,\ldots,g_n] \equiv
\left(K_i g_i\right)^{\frac{D}{D-2\Delta_i}} \,
{\cal F}_i\left(\frac{K_j g_j}{(K_i g_i)^{\phi_{ji}}}\right)
\,\,\,.
\label{scalingfree}
\EN
The functions ${\cal F}_i$ are universal homogeneous scaling
functions of the ratios $\frac{K_j g_j}{(K_i g_i)^{\phi_{ji}}}$.

Let us consider now the definition of the thermodynamical quantities
related to the various derivatives of the free--energy. We will
adopt the notation $\langle ... \rangle_i$ to denote expectation
values computed in the off--critical theory obtained by keeping
(at the end) only the coupling constant $g_i$ different from zero.
The first quantities to consider are the vacuum expectation values
(VEV) of the fields $\varphi_j$ which can be parameterized as
\EQ
\langle \varphi_j \rangle_i = -\left.\frac{\partial f_i}{\partial g_j}
\right|_{g_l=0} \equiv B_{ji}
g_i^{\frac{2 \Delta_j}{D-2 \Delta_i}} \,\,\,,
\label{vacuumji}
\EN
with
\EQ
B_{ji} \sim K_j K_i^{\frac{2 \Delta_j}{D-2 \Delta_i}} \,\,\,.
\label{bji}
\EN
The above relations can be equivalently expressed as
\EQ
g_i = D_{ij} \left(\langle \varphi_j\rangle_i\right)^
{\frac{D-2 \Delta_i}{2 \Delta_j}} \,\,\,,
\label{magn}
\EN
with
\EQ
D_{ij} \sim \frac{1}{K_i K_j^{\frac{D-2 \Delta_i}{2 \Delta_j}}}
\,\,\,.
\label{dij}
\EN
The generalized susceptibilities of the model are defined by
\EQ
\hat \Gamma_{jk}^i =
\frac{\partial}{\partial g_k} \langle \varphi_j\rangle_i =\left.
-\frac{\partial^2 f_i}{\partial g_k \partial g_j}\right|_{g_l=0}
\,\,\,. \label{susc}
\EN
They are obviously symmetrical in the two lower indices. By
extracting their dependence on the coupling constant $g_i$,
they can be expressed as
\EQ
\hat \Gamma_{jk}^i = \Gamma_{jk}^i \,g_i^
{\frac{2 \Delta_j + 2 \Delta_k - D}
{D - 2 \Delta_i}} \,\,\, ,
\label{hatgammajki}
\EN
with
\EQ
\Gamma_{jk}^i \sim K_j K_k K_i^{\frac{2 \Delta_j + 2 \Delta_k - D}
{D - 2 \Delta_i}} \,\,\, .
\label{gammajki}
\EN

The various quantities obtained by taking the derivatives of the 
free--energy obviously contain metric factors (the quantities $K_i$) 
which make their values not universal. However, it is always possible 
to consider special combinations thereof in such a way to cancel out 
all metric factors. Here we propose the consideration of the following 
universal ratios
\EQ
(R_c)^i_{jk} = \frac{\Gamma_{ii}^i \Gamma_{jk}^i}{B_{ji} B_{ki}}
\,\,\, ;
\label{Rc}
\EN
\EQ
(R_{\chi})^i_j = \Gamma_{jj}^i D_{jj} B_{ji}^
{\frac{D-4 \Delta_j}{2 \Delta_j}}
\,\,\,;
\label{Rchi}
\EN
\EQ
R^i_{\xi} = \left(\Gamma_{ii}^i\right)^{1/D} \xi_i^0 \,\,\,;
\label{Rxi}
\EN
\EQ
(R_A)^i_j = \Gamma_{jj}^i \, D_{ii}^
{\frac{4 \Delta_j + 2 \Delta_i - 2 D}{D-2 \Delta_i}} \,
B_{ij}^{\frac{2\Delta_j -D}{\Delta_i}} \,\,\, ;
\label{RA}
\EN
\EQ
(Q_2)^i_{jk} = \frac{\Gamma^i_{jj}}{\Gamma^k_{jj}}
\left(\frac{\xi_k^0}{\xi_j^0}\right)^{D-4 \Delta_j} \,\,\,.
\label{Q2}
\EN
From their definition, these quantities are pure numbers attached 
to the universality classes and can be used to characterize them. 
In fact, the amplitude ratios are numbers which typically present
significant variations between different classes of universality, 
whereas the critical exponents usually assume small values which 
only vary by a small percent by changing the universality classes. 
Hence the universal ratios may be ideal marks of the 
critical scaling regime \cite{Privman}. It is also worth emphasizing 
that, from an experimental point of view, it should be simpler to 
measure universal amplitude ratios rather than critical exponents: 
to determine the former quantities one needs to perform 
several measurements at a single, fixed value of the coupling 
which drives the system away from criticality whereas to determine 
the latter, one needs to make measurements over several decades 
along the axes of the off--critical couplings. Moreover, 
although not all of them are independent, the universal ratios 
are a larger set of numbers than the critical exponents and 
therefore permit a more precise determination of the class of
universality.

\section{Tricritical Ising Model}

The universal ratios of the two--dimensional Tricritical Ising 
Model (TIM) were computed in refs.\cite{FMS1,FMS2}. We refer the 
reader to these articles for a detailed discussion of their evaluation. 
There are several reasons for considering the class of universality 
associated to this model. First of all, from the experimental point of 
view, a number of physical systems exhibit a tricritical Ising behavior, 
among them fluid mixtures, metamagnets or order--disorder transitions in
absorbed systems \cite{Tejwani} (for a review on the theory of 
tricritical points, see \cite{Lawrie}). Secondly, from a 
theoretical point of view, the TIM is still sufficiently
simple to be solved but at the same time it presents an extremely 
rich and fascinating structure of excitations away from criticality.  
In a Landau--Ginzburg approach, the TIM is associated to a 
$\Phi^6$--theory near its tricritical point \cite{ZamLG}, 
although this approach is often too elementary for the
understanding of some of its remarkable symmetries. Depending 
on the direction in the phase space in which the system is 
moved away from criticality, one can observe, for instance, 
a behavior ruled by the exceptional root system $E_7$ \cite{MC,FZ} 
or by supersymmetry \cite{Kastor,Zammassless,DMSmassless,Zamthree} 
(in its exact or broken phase realization) or by an asymmetrical 
pair of kinks \cite{LMC,Smirnov12,CKM}. In addition, the description
of its low--temperature phase is easily obtained from the one of
its high--temperature phase because of the self--duality of
the model. 

A convenient way to determine the universal ratios of the 
TIM consists of adopting a Quantum Field Theory (QFT) approach, 
as it was done in the original references \cite{FMS1,FMS2}. It would 
be obviously interesting to compute and to compare them by 
using a lattice formulation as, for instance, the one studied in 
\cite{dilute}. In a QFT approach, one takes initially 
advantage of the exact solution of the model at criticality. 
The bidimensional TIM is described by the second representative
of the unitary series of minimal models of CFT \cite{BPZ,FQS1}:
its central charge is equal to $c = \frac{7}{10}$ and the 
exact conformal weights of the scaling fields are given 
by 
\EQ
\Delta_{l,k} = \frac{(5 l - 4 k)^2 -1}{80} \,\,\,\,\,\,\,,
\,\,\,\,\,
\begin{array}{c}
1 \leq l \leq 3 \\ 1 \leq k \leq 4
\end{array}
\label{Kactable}
\EN
There are six primary scalar fields $\phi_{\Delta,\overline\Delta}$, 
which close an algebra under
the Operator Product Expansion
\begin{eqnarray}
&& \phi_i(z_1,\overline z_1) \,
\phi_j(z_2,\overline z_2) \,  \sim \, \\
&& \sim \sum_k \, c_{ijk} \mid z_1 - z_2
\mid^{-2 (\Delta_i + \Delta_j - \Delta_k)} \phi_k(z_2,\overline z_2)
\,\,\,.\nonumber 
\end{eqnarray}
The skeleton form of this OPE algebra and the relative structure
constants of the Fusion Rules of the TIM were given in ref.\,
\cite{LMC}. The six primary fields can be identified with the 
normal ordered composite LG fields \cite{ZamLG}. With
respect to their properties under the $Z_2$ spin--reversal 
transformation $Q: \Phi \rightarrow - \Phi$ we have:
\begin{enumerate}
\item two odd fields: the leading magnetization operator
$\sigma = \phi_{\frac{3}{80},\frac{3}{80}} \equiv \Phi$
and the subleading magnetization operator $\sigma' =
\phi_{\frac{7}{16},\frac{7}{16}} \equiv : \Phi^3:$
\item four even fields: the identity operator $1 = \phi_{0,0}$,
the leading energy density $\varepsilon = \phi_{\frac{1}{10},
\frac{1}{10}} \equiv :\Phi^2:$, the subleading energy density
$t = \phi_{\frac{6}{10},\frac{6}{10}} \equiv :\Phi^4: $, which in
metamagnets assumes the meaning of the density of the
annealed vacancies, and the field $\varepsilon" = \phi_{\frac{3}{2},
\frac{3}{2}}$. The OPE of the even fields form a subalgebra of
the Fusion Rules.
\end{enumerate}
In the TIM there is another $Z_2$ transformation -- the
Kramers--Wannier duality $ D$ -- under which the fields
transform as follows:
\begin{itemize}
\item          
the order magnetization operators are mapped onto their
corresponding disorder operators
\EQ
\mu =  D^{-1} \sigma  D
\,\,\,\,\, ,
\,\,\,\,\,
\mu' =  D^{-1} \sigma' D 
\,\,\,. \label{sigmadual}
\EN
\item
the even fields are mapped onto themselves,
\EQ
D^{-1} \varepsilon D =-\varepsilon \hspace{3mm},
\hspace{3mm}
D^{-1} t D = t  \hspace{3mm}, 
\EN   
\end{itemize}
Given the $Z_2$--spin odd parity of the two magnetic fields, 
the off--critical theories obtained by their perturbation 
are independent of the sign of their associate couplings. 
The odd parity of the energy field under the other $Z_2$ 
symmetry is responsable for the duality mapping between the 
high and low temperature phases. On the other hand, two 
different physical situations arise depending on the sign 
of the coupling relative to the perturbation of the 
vacancy density field $t(x)$. It is important to notice 
that, excluding the leading magnetic perturbation, all the 
others give rise to integrable QFT away from criticality. 
To simplify the formulae below, it is convenient
to adopt the compact notation $\varphi_i$ $(i=1,2,3,4)$ to
denote collectively all these fields, so that $\varphi_1
= \sigma$ , $\varphi_2 = \varepsilon$, $\varphi_3=\sigma'$
and $\varphi_4=t$. Let us discuss now some aspects of QFT 
which make possible the determination of the universal 
ratios and the strategy we have used to achieve these results. 

\section{Quantum Field Theory Approach}

Each coupling constant $g_i$ ($i=1,\ldots,4$) relative 
to the relevant operator $\varphi_i(x)$ of the TIM 
is a dimensional quantity which can be related to the lowest 
mass--gap $m_i = \xi_i^{-1}$ of the off--critical theory 
according to the formula 
\EQ
m_i = {\cal C}_i \,g_i^{\frac{1}{2-2 \Delta_i}} \,\,\, .  
\label{mg}
\EN 
When the QFT associated to the action (\ref{actionoff}) 
is integrable, the pure number ${\cal C}_i$ can be exactly 
determined by means of the Thermodynamical Bethe Ansatz 
\cite{TBA,fateev}. When the theory is not integrable (this is 
the case for the magnetic deformation of the TIM), the constant 
${\cal C}_i$ can be nevertheless determined by a numerical method, 
based on the so--called Truncated Conformal Space Approach 
\cite{YZ}. In conclusion, for all individual deformations of 
the TIM we are able to completely set the relationship which 
links the coupling constant to the mass--gap of the theory and 
therefore to switch freely between these two variables. 

Another set of quantities which can be fixed by QFT are the matrix 
elements of the order parameters, the simplest ones being the vacuum 
expectation values (VEV). In this case we have 
\EQ
\langle \varphi_j \rangle_i = B_{ji} \,g_i^{\frac{\Delta_j}
{1- \Delta_i}} 
\,\,\,.
\label{VEVg}
\EN 
When the theory is integrable, the constant $B_{ji}$ can 
be fixed exactly, thanks to the results of a remarkable series of 
papers \cite{russian1,russian2}. When it is not integrable, the 
constant $B_{ji}$ can be nevertheless estimated by means 
of a numerical approach, as firstly shown in \cite{GM1}. 
Hence, also in this case, we are able to determine completely 
these quantities. Moreover, as shown in \cite{FMS2}, 
a generalization of the numerical approach of ref.\,\cite{GM1} 
often leads to a reasonable estimation of the matrix elements of 
the order parameters between the vacuum states and some of the 
excited states, as for instance 
$
\langle 0 | \varphi_j | A_k \rangle_i 
$ 
where $A_k$ is a one--particle state of mass $M_k$. These 
quantities turn out to be useful for obtaining sensible
approximation of the large--distance behavior of several 
correlators. 

Another useful piece of information on the off--critical dynamics 
can be obtained by exploiting the properties of the stress--energy 
tensor $T_{\mu \nu}(x)$. In the presence of the perturbing field 
$\varphi_i$, the trace of the stress--energy tensor is different 
from zero and can be expressed as 
\EQ
\Theta(x) = 2 \pi g_i (2 - 2 \Delta_i) \,\varphi_i \,\,\,.
\label{trace}
\EN 
The trace of the stress--energy tensor enters a useful sum 
rule -- called the $\Delta$--theorem sum rule \cite{DSC} -- 
which reads 
\EQ
\Delta_j = -\frac{1}{4 \pi \langle \varphi_j \rangle_i} 
\, \int d^2x \,\langle \Theta(x) \,\varphi_j(0) \rangle_i^c 
\,\,\,, 
\label{Deltath}
\EN 
{\it i.e.} it relates the conformal dimension $\Delta_j$ of the 
field $\varphi_j$ to its VEV and to the integral of its connected 
off--critical correlator with $\Theta(x)$. It is easy to 
see that the above formula simply expresses the content of the 
fluctuation--dissipation theorem and when the above integral 
diverges, so does the VEV in the denominator, in such a way 
that eq.\,(\ref{Deltath}) always keeps its validity \cite{FMS2}.  

Basic quantities in the universal ratios are the generalized 
susceptibilities $\Gamma_{jk}^i$ which, by using the 
fluctuation--dissipation theorem, can be expressed as 
\EQ
\hat\Gamma_{jk}^i = \int d^2 x \langle \varphi_j(x) 
\varphi_k(0)\rangle_{c}^i 
\,\,\,.
\label{fluctdis}
\EN 
By extracting their dependence on the coupling constant $g_i$, 
we have $\hat \Gamma_{jk}^i = 
\Gamma_{jk}^i\, g_i^{\frac{\Delta_j + \Delta_k-1}{1-\Delta_i}}$ 
with 
\EQ
\Gamma_{jk}^i = {\cal C}_i^{2 \Delta_j+2\Delta_k-2} 
\int d\tau \frac{1}{\tau^{2\Delta_j+2\Delta_k}} Q_{jk}(\tau) 
\,\,\,.
\label{finalgamma}
\EN
Some of the above susceptibilities can be determined exactly, 
such as the components $\Gamma_{ik}^i$, whose values are 
provided by the $\Delta$--theorem sum rule 
\EQ
\Gamma_{ik}^i = - \frac{\Delta_k}{1-\Delta_k} B_{ki} \,\,\,.
\label{gammachiusa}
\EN 
In all other cases, when an exact formula is not available, 
the strategy to evaluate the generalized susceptibilities  
relies on two different representations of the correlators. 
These representations have the advantage to converge
very fast in two distinct regions. 

The first representation is based on Conformal Perturbation 
Theory (with the employment of the non--analytic expression 
of the VEV) \cite{ZamYL}. In this approach the two--point
correlators are expressed as 
\EQ
\langle\varphi_i(x) \varphi_j(0)\rangle = 
\sum_i C_{ij}^p(g;x) \langle A_p(0)\rangle  \,\,\,
\label{correlator}
\EN
where the off--critical structure constants 
$C_{jk}^p(g;x)$ admit the 
expansion
\EQ
C_{ij}^p(g;x)= r^{2(\Delta_p-\Delta_i-\Delta_j)} \,
\sum_{n=0}^{\infty}C_{i,j}^{p(n)}(gr^{2-2\Delta_{\Phi}})^n 
\,\,\, ,
\label{Cexpansion}
\EN
($r=\mid x\mid$) and can be computed perturbatively in $g$. 
Their first order contribution is given by \cite{ZamYL}
\EQ
C_{i,j}^{p(1)} = -\int ' d^2w \,\langle \tilde{A}^p(\infty) 
\tilde{\Phi}(w)\tilde{\varphi}_i(1)\tilde{\varphi}_j(0)\rangle_{CFT} 
\,\,\,,
\label{firstorderc}
\EN
where the prime indicates a suitable infrared (large distance) 
regularization of the integral. This representation allows a 
very efficient estimation of the correlation function in its 
short distance regime $r \ll \xi$. 

The second representation is based on the Form Factors and allows 
an efficient control of its large distance behavior, i.e. 
when $r \gg \xi$. In this second representation, one makes
use of the knowledge of the off--critical mass spectrum of 
the theory to express the correlators 
as
\EQ
\langle\varphi_i(x) \varphi_j(0)\rangle = \sum_{n=0}^{\infty} 
g_n(r) \,\,\,,
\label{g_n}
\EN 
where 
\begin{eqnarray}
\label{defff}
&& g_n(r)  =  \int_{\theta_1 >\theta_2 \ldots>\theta_n} 
\frac{d\theta_1}{2\pi} \dots \frac{d\theta_n}{2\pi}\,
\,e^{-r \sum_{k=1}^n m_k \cosh\theta_k} 
\nonumber \label{spectral} \\
&& 
\times \langle 0|\varphi_i(0)|\dots A_{a_n}(\theta_n)
\rangle  
\langle \dots A_{a_n}(\theta_n)|
\varphi_j(0)|0 \rangle 
\,\,\,. 
\nonumber
\end{eqnarray} 
$|A_{a_1}(\theta_1) \dots A_{a_n}(\theta_n)\rangle$ are the 
multi--particle states relative to the excitations of mass 
$m_k$, with relativistic dispersion relations given by $E 
= m_k \cosh \theta$, $p = m_k \sinh\theta$, where $\theta$ 
is the rapidity variable. The spectral representation 
(\ref{spectral}) is obviously an expansion in the parameter 
$e^{-\frac{r}{\xi}}$, where $\xi^{-1} = m_1$ is the lowest 
mass--gap. 

Basic quantities of the large distance approach are the Form 
Factors (FF), {\it i.e.} the matrix elements of the operators 
$\varphi_i$ on the physical asymptotic states \cite{KW,Smirnov}
\EQ
F^{\varphi_i}_{a_1,\ldots ,a_n}(\theta_1,\ldots,\theta_n) \,= 
\, \langle 0| \varphi_i(0)|A_{a_1}(\theta_1)\ldots A_{a_n}
(\theta_n) \rangle \,\,\,.
\label{form}
\EN
It is worth emphasizing that the above quantities are unaffected 
by renormalization effects since physical excitations are employed 
in their definitions. For scalar operators, relativistic invariance 
requires that the FF only depend on the rapidity differences 
$\theta_i - \theta_j$. In our calculation we have only 
inserted into the spectral representations the one--particle 
and two--particle FF, computed according to the analysis 
of ref.\,\cite{DMIMMF,AMV}. 

Both the representations (\ref{correlator}) and (\ref{g_n}) 
are known to converge very fast (see, for instance \cite{CMpol,ZamYL}
and therefore they are efficiently approximated by their lowest terms, 
which therefore can be evaluated with a relatively little analytical 
effort. These  considerations obviously lead to the estimation of the 
integral (\ref{fluctdis}) according to the following steps: 
\begin{enumerate}
\item Express the integral in polar coordinates as 
\EQ
\hat\Gamma_{jk}^i = 2\pi \int_0^{+\infty} 
d r \,r \,\langle\varphi_j(r) 
\varphi_k(0)\rangle_c^i \,\,\, , 
\EN 
and split the radial integral into two pieces as 
\begin{eqnarray}
& & I  = \int_0^{+\infty}  
d r \,r\,\langle \varphi_j(r) 
\varphi_k(0)\rangle_c^i 
\nonumber \\ 
&& =  \int_0^{R}  
d r \,r\,\langle
\ldots 
\rangle_c^i +  
\int_R^{+\infty}  
d r \,r\,\langle\ldots 
\rangle_c^i \nonumber 
\\
& & \equiv I_1(R) + I_2(R) 
\label{I1I2}
\,\,\,.
\end{eqnarray}
\item 
Use the best available short--distance representation of the 
correlator to evaluate $I_1(R)$ as well as the best available 
estimate of its large--distance representation to evaluate 
$I_2(R)$. 
\item Optimize the choice of the parameter $R$ in such a 
way to obtain the best evaluation of the whole integral. 
In practice, this means looking at that value of $R$ 
for which a plateau is obtained for the sum of $I_1(R)$ 
and $I_2(R)$. Say in another way, $R$ belongs to that 
interval where there is an overlap between the short--distance 
and the long--distance expansion of the correlator (see, 
for instance, Figure 1, relative to the correlator 
$\langle \varphi_1(x) \varphi_1(0)\rangle_1$ in the
high--temperature phase). 
\end{enumerate}
\begin{figure}[h]
\begin{center}
\psfig{figure=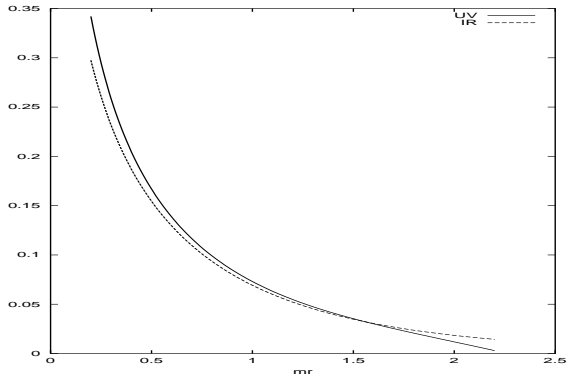,height=5cm,width=8cm}
\caption{
Continuous line = UV approximation, dashed line = 
IR approximation. An overlap of the curves is observed
around $mr\sim 1$.
}
\label{cor112}
\end{center}
\end{figure}  
Gathering all the results relative to the above quantities, 
a set of universal ratios for the TIM have been obtained. Some 
of them are exact, like $(R_c)^{1}_{1,k} = \frac{240}{5929} 
\Delta_k$, $(R_c)^2_{2,k} =\frac{10}{81} \Delta_k$ ($k=1,\ldots,4$). 
Those relative to the low and high temperature phase of the model 
denoted by an upper index $\mp$ respectively, are 
in Table 1. An interesting universal ratio is provided in this case 
by the correlation lengths $\xi^{\pm}$, measured at the same 
displacement above and below the critical temperature 
(as extracted from the correlation function of the magnetic operator 
using its duality properties)
\EQ
{\xi^+\over \xi^-} = 2\cos\left({5\pi\over 18}\right) \approx 
1.28557...
\label{ratxi0}
\EN
which can be inferred by the exact mass spectrum of the model and the 
parity properties of the excitations \cite{MC,LMC}. Other universal 
ratios are presented in Tables 2--5 and a more complete set of them 
can be found in ref.\cite{FMS2}.

In view of the predictivity showed by the theoretical approach, 
it would be interesting to have their experimental confirmation. 
It would be equally useful to apply the methods discussed here 
to other models in such a way to bridge a closer contact between 
theoretical and experimental results in two--dimensional physics. 

\acknowledgments
This work was supported by the TMR Network "Integrability, 
non--perturbative Effects and Symmetry in Quantum Field Theory', 
contract number FMRX-CT96-0012.

\vspace{15mm}

\noindent {\bf Table 1:} Amplitude ratios 
$R^2_{jk}={\Gamma_{jk}^{2+}\over \Gamma_{jk}^{2-}}$.
\vskip 3mm
\begin{center}
\begin{tabular}{|ccc||ccc|}
\hline
$R^2_{11} $   &  = &$3.54$  &
$R^2_{13} $   &  = &$-2.06 $  \\
\hline
$R_{22}^2 $   &  = &$1$  &
$R_{24}^2 $   &  = &$-1$  \\
\hline
$R_{33}^2 $   & = &$ 1.30$  &
$R_{44}^2$   & = &$1$  \\
\hline
\end{tabular}
\end{center}

\noindent {\bf Table 2:} Universal ratios $(R_c)_{jk}^{1}$ and
$(R_c)_{jk}^{2-}$.
\vskip 3mm
\begin{center}
\begin{tabular}{|ccl||ccl|} \hline
$(R_c)_{22}^1$   & = & $ 1.05~10^{-2}$   &  $(R_c)_{23}^1$    
& = & $4.85~10^{-2}$  \\
\hline
$(R_c)_{24}^1$   & = & $ 6.7~10^{-2} $   &  $(R_c)_{33}^1$    
& = & $3.8~10^{-1} $ \\
\hline \hline
$(R_c)_{11}^{2-}$ & = &$ 2.0~10^{-3}$  & $(R_c)_{14}^{2-}$ & = & 
$-2.34~10^{-2}$  \\
\hline
$(R_c)_{13}^{2-}$ & = &$ 1.79~10^{-2}$ & $(R_c)_{33}^{2-}$ &  = & 
$3.4~10^{-1}$  \\
\hline
\end{tabular}
\end{center}

\noindent {\bf Table 3:} Universal ratio $(R_{\chi})_{j}^{i}$ for
$i,j=1,2$.
\vskip 3mm
\begin{center}
\begin{tabular}{|ccl||ccl|} \hline
$(R_{\chi})_{1}^1$   & = &  $3.897~10^{-2}$ & $(R_{\chi})_2^{2+} 
$  & = & $0.1111$  \\
\hline
$(R_{\chi})_{2}^1$   & = &  $0.116$         & $(R_{\chi})_{1}^{2-} 
$  & = & $0.040$  \\
\hline
$(R_{\chi})_{1}^{2+}$ & = & $ 0 $           & $(R_{\chi})_{2}^{2-} 
$ & =  & $0.1111$  \\
\hline
\end{tabular}
\end{center}

\noindent {\bf Table 4:} Universal ratios $R_{\xi}^i$ and 
$(R_A)_{j}^{i}$ for $i,j=1,2^-,2^+$.
\vskip 3mm
\begin{center}
\begin{tabular}{|ccl||ccl|} \hline
$R_{\xi}^1$      & = & $ 7.557~10^{-2}$     &  & & \\
\hline
$R_{\xi}^{2+}$   & = & $ 1.0784~10^{-1} $   &  $R_{\xi}^{2-}$    
& = & $8.389~10^{-1} $ \\
\hline \hline
$(R_A)_{2+}^{1}$ & = & $0 $  & $(R_A)_{2-}^{1}$ & = & 
$3.918~10^{-2}$  \\
\hline
$(R_A)^{2+}_{1}$ & = & $ 2.958~10^{-1}$ & $(R_A)^{2-}_{1}$ &  = & 
$8.260~10^{-1}$  \\
\hline
\end{tabular}
\end{center}

\noindent {\bf Table 5:} Universal ratios $(Q_2)^i_{jk}$ for
$i,j,k=1,2^+,2^-$.
\vskip 3mm
\begin{center}
\begin{tabular}{|ccl||ccl|} \hline
$(Q_2)^1_{2^+1}$   & = & $ 1.260$     & $(Q_2)^1_{2^-1}$  
& = &  $1.884$ \\
\hline
$(Q_2)^1_{2^+2^+}$ & = & $ 1.973$     & $(Q_2)^1_{2^+2^-}$     
& = &  $1.320$ \\
\hline 
$(Q_2)^{2+}_{11}$  & = & $ 1.56 $    & $(Q_2)^{2-}_{11}$ 
& = &  $0.442$  \\
\hline
$(Q_2)^{2+}_{12^-}$ & = & $ 1.70$     &  &   & \\
\hline
\end{tabular}
\end{center}

\end{document}